\documentclass[a4paper]{jpconf}
\usepackage{graphicx}
\begin{document}
\title{Linear Solar Models: a simple tool to investigate the properties of solar interior}

\author{F.L. Villante}

\address{Universit\`a di L'Aquila and INFN-LNGS, L'Aquila, Italy}

\ead{villante@lngs.infn.it}

\begin{abstract}
We describe a simple method to study the dependence of the solar properties on
a generic (small) modification the physical inputs adopted in standard solar models calculations.
\end{abstract}

\section{Introduction}

 A new solar problem has emerged recently. The determinations of the photospheric
heavy element abundances \cite{asplund} indicate that the sun metallicity is lower than previously 
assumed \cite{gs98}. Solar models that incorporate these lower abundances are no more able to 
reproduce the helioseismic results. 
Detailed studies have been done to resolve this 
controversy (see \cite{HelioReview} for a review), but the origin
of the ``solar composition problem''  has still to be identified. 

In this situation, it is useful to perform a detailed analysis of the role of physical inputs and of 
the standard assumptions for standard solar model (SSM) calculations. This task is not always possible in simple terms,
since the construction of SSM relys on (time-consuming) numerical integration of a non linear system 
of partial differential equations. 
In order to overcome this problem, we provide a tool which is at same time simple and accurate enough to
describe the effects of a generic (small) modification of the physical inputs on the solar properties .
The starting point is the idea that, despite the present disagreement with helioseismic data, the 
SSM is a rather good approximation of the real sun.  
We can thus assume small variations of the physical and chemical properties of the sun with respect to the SSM predictions 
and use a linear theory to relate them to the properties of the solar plasma.

\section{Linear Solar Models}

We consider the effects of a generic (small) modification of the input parameters of the sun.
Namely, we indicate with $\delta \kappa(r)$ and $\delta \epsilon(r)$ 
the relative variations of radiative opacity $\kappa(r)$ and of the energy generation 
$\epsilon(r)$ {\em along the SSM profile}, i.e.: 
\begin{equation}
I(\overline{T}(r),\overline{\rho}(r),\overline{Y}(r),\overline{X}_{i}(r))=
\overline{I}(\overline{T}(r),\overline{\rho}(r),\overline{Y}(r),\overline{X}_{i}(r))\;[1+\delta I(r)]
\end{equation} 
where $I=\kappa,\,\epsilon$ and $r$ indicates the distance from the center of the sun. 
The notation $\overline{Q}$ indicates, here and in the following, the SSM value for the generic quantity $Q$.
As a result of this modification, we obtain a solar model which deviates from SSM. 
We indicate with $\delta h(r)$ the relative variation of the physical quantities $h = l,\,m,\,T,\,P$ with 
respect to SSM predictions, i.e.: 
\begin{equation}
\nonumber
h(r) = \overline{h}(r)\,
[1+\delta h(r)]\\
\end{equation}
%
 In the radiative region of the sun ($r\le \overline{R}_{\rm b}=0.730 R_{\odot}$), the functions $\delta h(r) $ are described 
with good accuracy by the solutions of the linear system of ordinary differential equations:
\begin{eqnarray}
\frac{d\delta m}{dr} &=& \frac{1}{l_m} \,\left[ \gamma_{P}\, \delta P + \gamma_T \, \delta T - \delta m 
+ \gamma_Y \, \Delta Y_{\rm ini}  + \gamma_\epsilon \; \delta \epsilon \right]\label{linsyst3}\\
\nonumber
\frac{d\delta P}{dr}&=& \frac{1}{l_P}\, \left[\left(\gamma_{P}-1\right) \, \delta P + \gamma_T \, \delta T + \delta m  
+ \gamma_Y \, \Delta Y_{\rm ini}  + \gamma_{\epsilon} \; \delta \epsilon \right]\\
\nonumber
\frac{d\delta l}{dr}& = & \frac{1}{l_l}\, \left[ \beta'_P \,\delta P +  \beta'_T \,\delta T - \delta l + 
 \beta'_Y \,\Delta Y_{\rm ini} + \beta'_C \, \delta C + \beta'_\epsilon \,\delta \epsilon \right]\\
\nonumber
\frac{d\delta T}{dr}&=& \frac{1}{l_T}\, \left[ \alpha'_P \,\delta P + \alpha'_T \,\delta T + \delta l + 
\alpha'_Y \, \Delta Y_{\rm ini}+ \alpha'_C \, \delta C + \delta \kappa  + \alpha'_\epsilon \, \delta \epsilon \right] 
\label{linear}
\end{eqnarray}
The coefficients $\gamma_{h}$, $\beta'_{h}$ and $\alpha'_{h}$ and the scale heights $l_{h}\equiv \left[d\ln( \overline{h})/dr\right]^{-1} $ have been
calculated in \cite{noi} and are shown in fig.1 and fig.2 (left panel). The parameters $\Delta Y_{\rm ini}$ and $\delta C$ represent 
the absolute variation of the initial helium abundance and the relative variation of pressure at the
bottom of the convective enevelope and can be univocally determined by imposing the appropriate integration conditions.  
At the center of the sun ($r=0$) we have:
\begin{eqnarray}
\nonumber
\delta m &=& \gamma_{P,0}\, \delta P_{0} + \gamma_{T,0} \, \delta T_{0} 
+ \gamma_{Y,0} \, \Delta Y_{\rm ini}  + \gamma_{\epsilon,0} \; \delta \epsilon_{0} \;\;\;\;\;\;\;\; \;\;\;\;\;\;\;\;\;\;\;\;\;\;\;\;\;\;\;\;
\delta P = \delta P_{0}\\
\nonumber
\delta l &=& \beta'_{P,0} \,\delta P_{0} +  \beta'_{T,0} \,\delta T_{0} + 
 \beta'_{Y,0} \,\Delta Y_{\rm ini} + \beta'_{C,0} \, \delta C + \beta'_{\epsilon,0} \,\delta \epsilon_{0}  \;\;\;\;\;\;\;\; \;\;\;\;\;\;\;\;\;\;\;\;\;
\delta T = \delta T_{0} 
\end{eqnarray}
where the subscript "0" indicates that a given quantity is evaluated at $r=0$. At the bottom of the convective envelope ($r=\overline{R}_{\rm b}$), 
we have instead:
\begin{eqnarray}
\nonumber
\delta m &=& - m_{\rm c} \, \delta C \;\;\;\;\;\;\;\; \;\;\;\;\;\;\;\;\;\;\;\;\;\;\;\;\;\;\;\;\;\;
\delta P  =  \delta C\\
\nonumber
\delta l &=& 0      \;\;\;\;\;\;\;\;\;\;\;\;\;\;\;\;\;\;\;\;\;\;\;\;\;\;\;\;\;\;\;\;\;\;\;\;\;\;
\delta T = A'_Y \, \Delta Y_{\rm ini} + A'_{C} \, \delta C
\end{eqnarray}
where $A'_Y=0.626$ and $A'_C=0.025$ and $m_{\rm c}=M_{\rm conv}/M_{\odot}=0.0192$ is the fraction of solar mass contained in the convective region.


\begin{figure}[t]
\begin{center}
\includegraphics[width=5cm,angle=0]{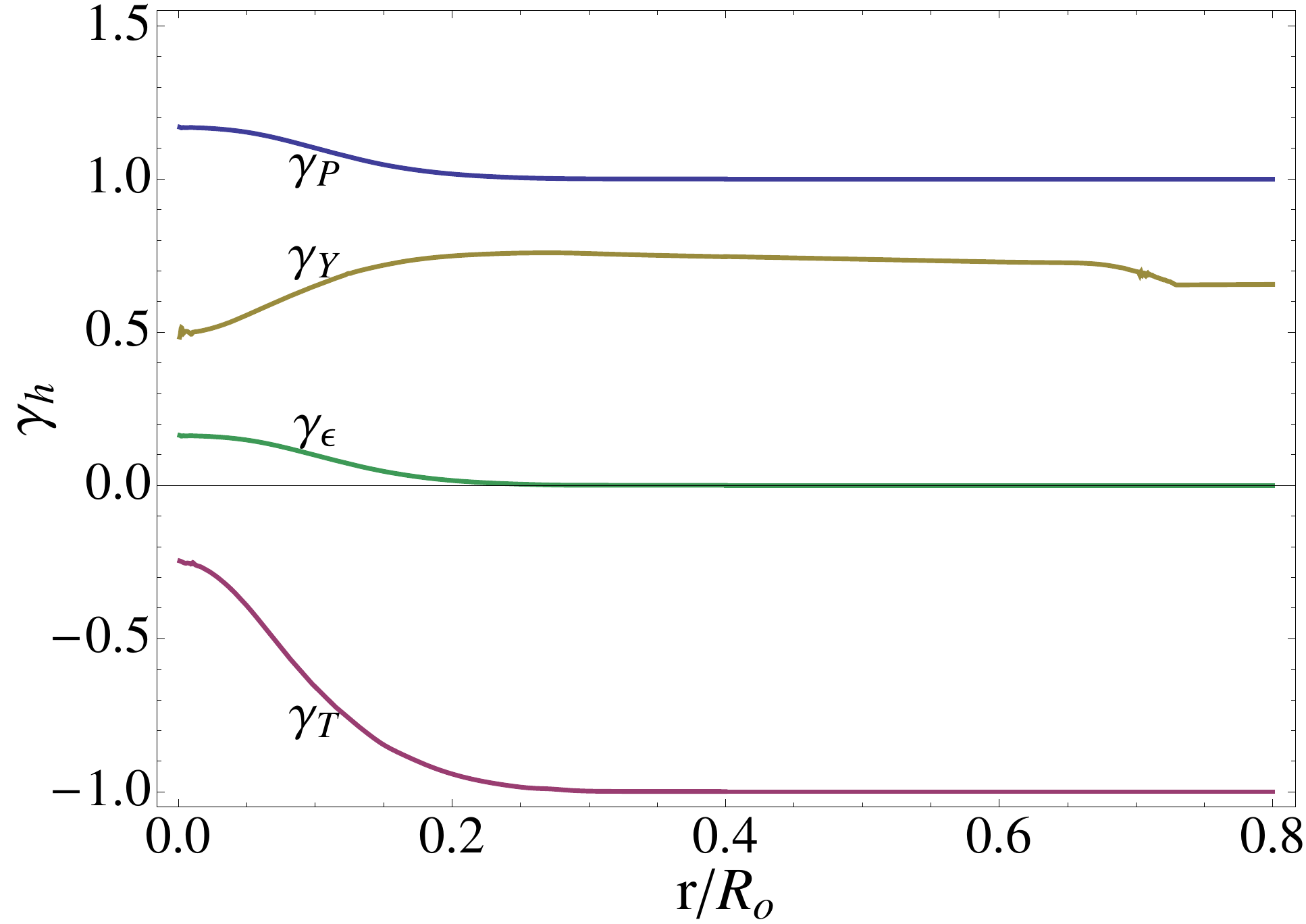}
\includegraphics[width=5cm,angle=0]{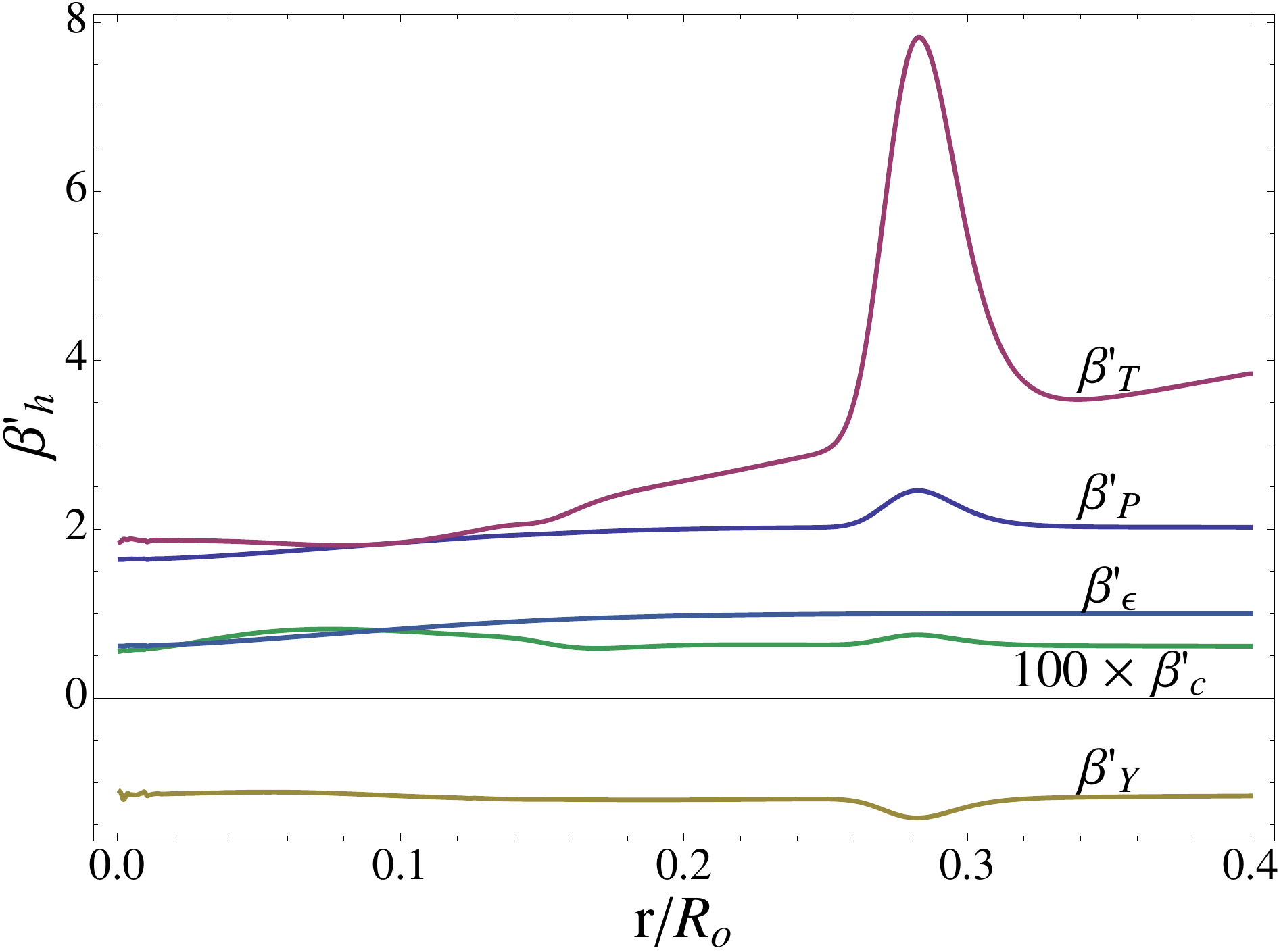}
\includegraphics[width=5cm,angle=0]{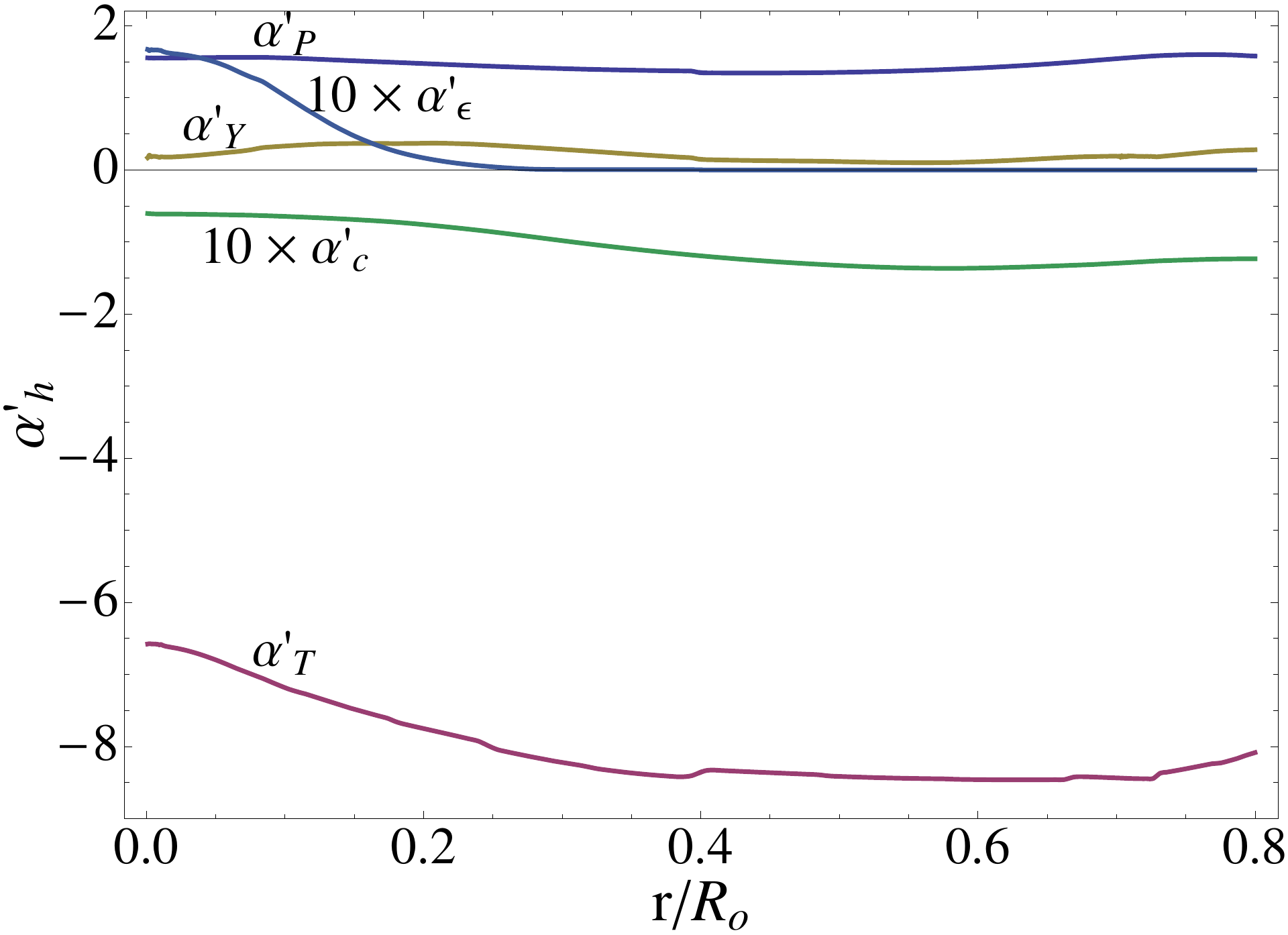}
\end{center}
\vspace{-5mm} \caption{ {\protect\small 
The coefficients $\gamma_{h}(r)$ (left panel), $\alpha'_{h}(r)$ (middle) and $\beta'_{h}(r)$ (right). See \cite{noi} for details.}}
\label{Fig2}
\end{figure}


The above equations completely determine the physical and chemical properties of the ``perturbed'' solar models.
They have been obtained in \cite{noi} by expanding to linear order the structure equations of the present sun close 
to the SSM solution.
We, thus, refer to the models calculated by using this approach as Linear Solar Models (LSM).
In the derivation of eqs.(\ref{linear}) we also assumed that:
\begin{enumerate}
\item
The relative variation of density $\delta \rho(r)$ can be described by:
\begin{equation}
\delta \rho(r) = \delta P(r) -\delta T(r) - P_{Y} \Delta Y(r)
\end{equation}
where $P_{Y}(r) \simeq - \partial \ln \mu / \partial Y \simeq - 5/[8-5Y(r)]$ and 
$\mu$ represents the mean molecular weight.
\item
The helium produced by nuclear reactions, $Y_{\rm nuc}(r)$, is roughly proportional to the 
energy generation coefficient in the present sun , i.e. $Y_{\rm nuc}(r)\propto \epsilon(r)$.
In this hypothesis (and neglecting the variation of elemental diffusion efficiency in the radiative region),
the absolute variation of helium $\Delta Y(r)$ is given by: 
 \begin{equation}
\Delta Y(r)=\xi_Y(r)\,\Delta Y_{\rm ini}+\xi_p(r)\,\delta P(r)+\xi_T(r)\,\delta T(r)+\xi_{\epsilon}(r)\,\delta\epsilon(r)
\label{helium}
\end{equation}
The coefficients $\xi_{h}(r)$ have been calculated in \cite{noi} and are shown in fig.2 (right panel).
\item
The global effect of elemental diffusion (i.e. integrated over the sun history) on the solar
surface abundances is roughly proportional to  diffusion efficiency at the bottom of the convective envelope
 in the present sun. In this assumption, one has:
 \begin{equation}
\Delta Y_{\rm b}= 0.838 \, \Delta Y_{\rm ini}+ 0.033\,  \delta C
\label{helium}
\end{equation}
where the subscript "b" indicates that a given quantity is evaluated at the bottom of the 
convective region. By taking into account that ($Z_{\rm b}/X_{\rm b}$) is observationally fixed, we can also calculate 
the relative variation of metals in the convective region:
 \begin{equation}
\delta Z_{\rm b}= -1.088 \,\Delta Y_{\rm ini} -0.043\,\delta C
\end{equation}
and in the radiative region (where we neglect the variation of diffusion efficiency):
 \begin{equation}
\delta Z(r) \simeq \delta Z_{\rm ini} = -0.887 \, \Delta Y_{\rm ini} - 0.164 \, \delta C
\end{equation}
\item 
The relative variation $\delta R_{\rm b}$ of the convective radius $R_{\rm b}$ can be calculated ``a-posteriori''. By 
applying the Swartzchild criterion to the solution of eqs.(\ref{linear}) and by expanding to first order (see \cite{noi} for details),
we obtain:
\begin{equation}
\delta R_{\rm b}=   0.449 \, \Delta Y_{\rm ini}  - 0.117  \, \delta C - 0.085 \, \delta\kappa_{\rm b}
\end{equation}
\end{enumerate}

In ref.\cite{noi} the predictions of LSM for helioseismic observables and neutrinos fluxes have been compared 
with the results of full non-linear SSM calculations for several selected cases, showing that a very good agreement 
is achieved.


\begin{figure}[t]
\par
\begin{center}
\includegraphics[width=6cm,angle=0]{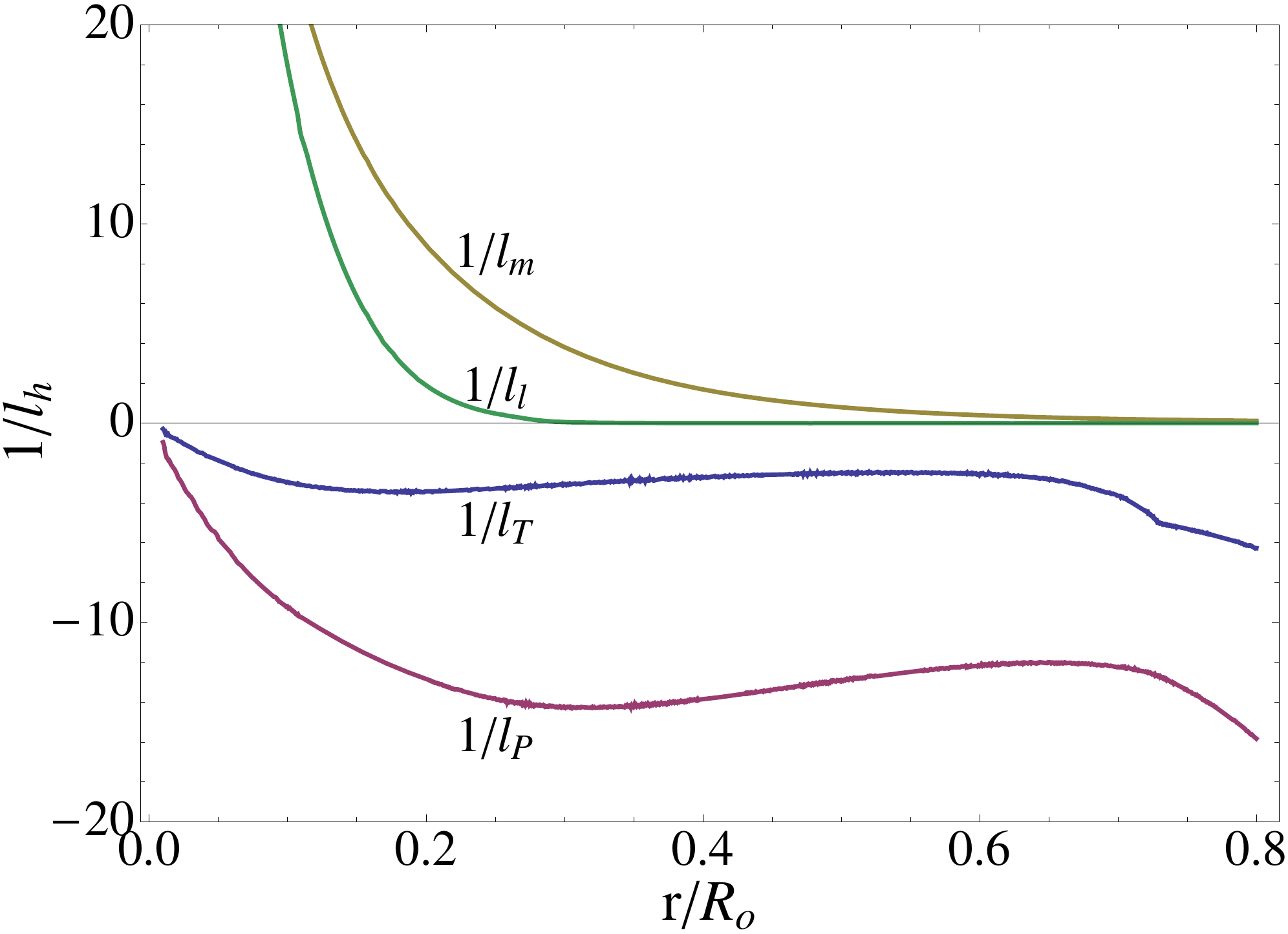}
\includegraphics[width=6cm,angle=0]{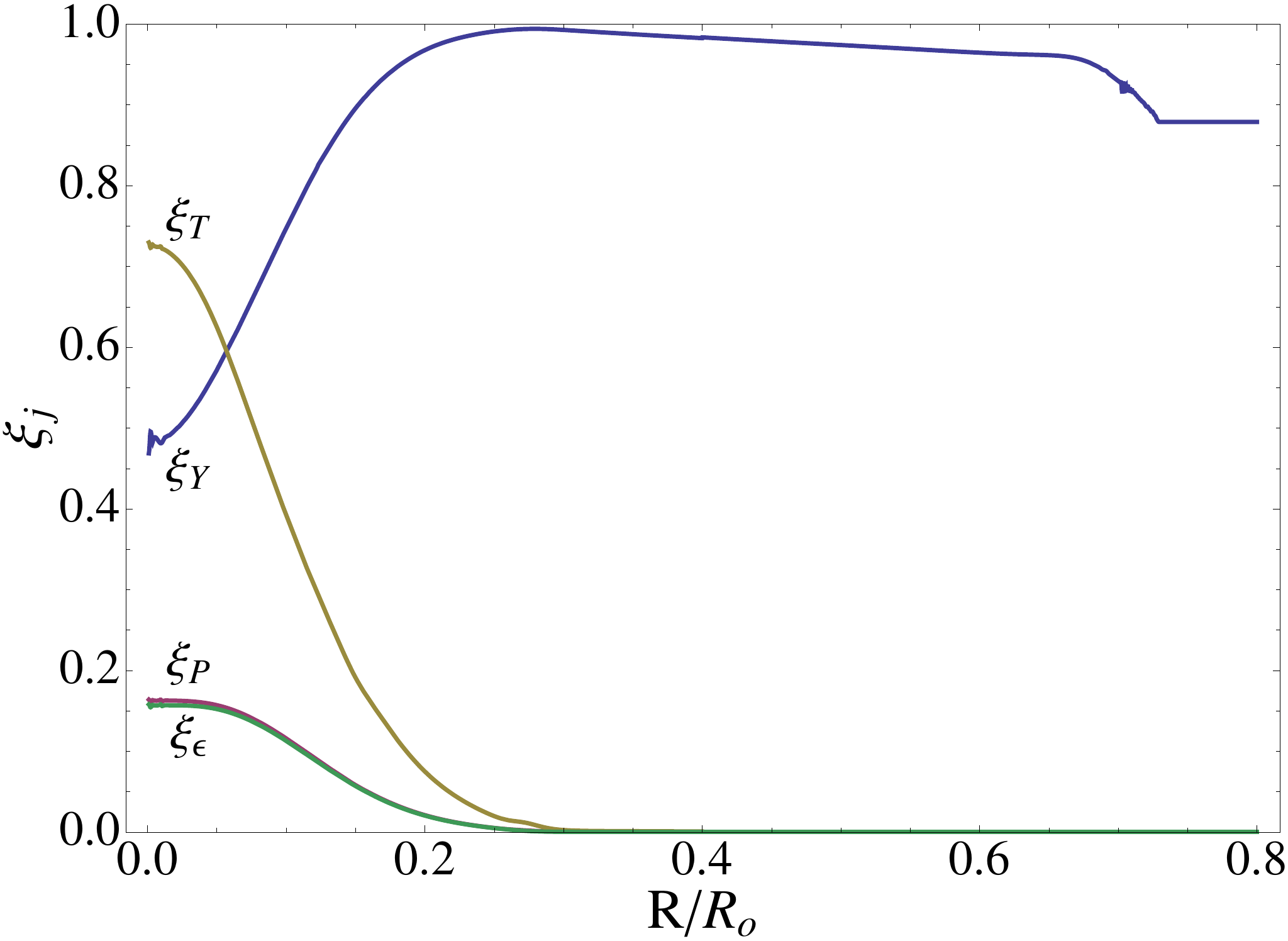}
\end{center}
\par
\vspace{-5mm} \caption{ {\protect\small The scale heights $l_{h}(r)$ (left panel) and the coefficients $\xi_{h}(r)$ (right panel). See \cite{noi} for details. }}
\label{Fig2}
\end{figure}


\end{document}